# WilloWISPs: A New Dark Growth Channel for Black Holes Suggests a Full-Spectrum Hierarchical MACHO Mass Function for Dark Matter

Zachary R. Smith and Neil F. Comins

Department of Physics and Astronomy, Bennett Hall, University of Maine, Orono, ME 04469

zachary.r.smith1@maine.edu,  galaxy@maine.edu

**Abstract:** Evidence of neutron stars with deconfined quark-matter cores suggests a new pathway for the evolution of black holes. New theories about the cores of neutron stars support the idea that quarkonium is likely to grow there as the neutron star ages. Surveys of stellar remnants have shown that there is no major mass gap between neutron stars and black holes. Black holes, specifically primordial ones (PBHs), have been suggested as an explanation for dark matter before. However, the way that very large black holes can form in the lifetime of the visible universe has only recently been explained by a promising solution to The Final Parsec Problem. If neutron stars can cool to become exotic stars or black holes surrounded by axions, then they may allow Intermediate-Mass Black Holes (IMBH) and Supermassive Black Holes (SMBH) to form quickly enough via coalescence. We find that a hierarchical clustering of Massive and Compact Halo Objects (MACHOs) with axion-dominated mini-halos can help to explain all of the missing dark matter. The model presented here suggests that this type of MACHO (WilloWISP) is likely equivalent to black holes above an unknown critical mass, which is less than ~ 1 $M_\odot$, and suggests that they ought to be quark stars below this mass. If quark stars are a transition state between neutron stars and black holes, then we find that black holes ought to be equivalent to boson stars, after all the residual quarkonium has formed a Bose-Einstein condensate of strange mesons.

## 1.    Introduction

Stellar evolution has been thoroughly explored, so the formation of hypothetical stars is difficult to place. Current theories and evidence about stellar remnants allow three options, depending on the mass of the progenitor star: white dwarf stars, neutron stars, and black holes. Both black holes and white dwarf stars are extremely stable. The same has been thought of neutron stars, which should only be able to become black holes if enough accretion occurs. However, exotic stars, like boson stars (Pombo & Saltas 2023; Chavanis & Harko 2012) and quark stars (Alford et al. 2006; Blaschke et al. 2001), could be a transition state between aging neutron stars and small black holes. Studies in the last few years support the existence of an exotic dark matter (DM) halo around each MACHO, which tidally dissipates orbital energy (Chan & Lee 2023; Noordhuis et al. 2023; Stelea et al. 2023). In this way, the many of the smallest MACHOs could be bridging the space with axions, which would drastically, and quiescently, increase the growth rate of larger black holes by coalescence.

Calculations of the speed of sound in neutron stars indicate that quarkonium, an extremely dense quark-gluon plasma (Ferrer 2017), is likely in the core of massive neutron stars (Annala 2000). Surveys of stellar remnants have shown that there is no major mass gap between neutron stars and black holes (Abbott et al. 2020). While some new work suggests that DM is not in the cores of neutron stars (Bell 2021), the statement refers to exotic DM, not the evolution of neutron stars into quark stars or black holes (Annala 2000).

We propose that quarkonium in the core could eventually deplete a neutron star's mantle until it is only a quark star, boson star, or a small black hole. These, along with several other hypothetical Massive and Compact Halo Objects (MACHOs) might, in fact, be of the same origins, or at least the same kind of core compositions (Pombo & Saltas 2023; Croker et al. 2020; Alford et al. 2006). Each possible MACHO may appear differently due to different aging conditions. The hypothesis that axions are collecting around the poles of pulsars (Noordhuis et al. 2023) is a sign that there is a strong-force mechanism at play that needs mediation of the weak force as a pulsar's jet is produced, and the star's mantle is consumed. The axion is an exotic particle proposed to solve the Strong-CP Problem (Charge and Parity violation related to the weak force from the exchange of color charges). In this work, we explore the possible connection between astrophysical jets;

black holes; hypothetical stars, including Proca (rotating boson) stars; pulsars (rotating neutron) stars; the strong force; axions; and DM.

Black holes, specifically primordial ones, have been suggested as an explanation for DM before, but only a small fraction (~15%) of the mass in a given galaxy can be attributed to them according to microlensing surveys (Rahvar 2005; Bird et al. 2023). Several mass gaps in the MACHO mass function have been probed, and Intermediate Mass Black Holes ($10^6$ M$_\odot$ ≳ IMBHs ≳ $10^3$ M$_\odot$) have become of particular interest in the last few years for their ability to explain a large fraction of the DM with a relatively small number of objects. Supermassive Black Holes (SMBHs > $10^6$ M$_\odot$) and IMBHs are thought to be far too rare and interactive with their environments to fit several observations, such as the anisotropies of the Cosmic Microwave Background (CMB) (Cann et al. 2018; Coriano & Frampton 2021; Hooper et al. 2024; Yang 2022). However, large MACHOs are potentially hiding in plain sight throughout galactic halos (Cann et al. 2018; Takekawa et al. 2019) since z ≳ 6 (Roberts et al. 2025; Wolf et al. 2018; Yue et al. 2024).

Constraints on the population of IMBHs (Bird et al. 2023) have been relaxed over the last few years due to bias against their detection via wide binaries (Yoo, Chaname & Gould 2004), supernovae lensing (García-Bellido, Clesse, & Fleury 2018), and x-ray binaries (Jonker et al. 2021). Furthermore, the CMB anisotropies do not account for direct-collapse, and other quiescent BH growth channels in the early universe, since bright tracers are necessary (Su, Li, & Feng 2023; Coriano & Frampton 2021; Frampton 2017). Dwarf galaxy dynamic heating is countered by axion halos around MACHOs, which cool their surroundings. Such quiescent IMBHs shrouded in cold gas and dust have been indirectly found in the Milky Way center (Takekawa et al. 2019), in the centers of dwarf galaxies (Almeida et al. 2024; Wang et al. 2023), and in the centers of some globular clusters (Häberle et al. 2024). The fact we have seen a "Nearly-Naked Rogue SMBH" (Condon 2017), or two, passing through galaxies also challenges the traditional idea that large black holes must only be the nuclei of large galaxies, as a result of galactic mergers.

The recently proposed solution to "The Final Parsec Problem" (Alonso-Álvarez, Cline, & Dewar 2024) by DM density spikes, and the increased quenching of star formation in dwarf galaxies via tidal shock (Hammer et al. 2024), further explain how black holes can reach large masses well within the age of the universe by coalescence and migration toward the center of stellar ensembles. The DM density spikes are thought to increase the orbital decay rate by tidally breaking (slowing) their companion stars' motion.

Gravitational waves are the primary means to dissipate kinetic energy in a binary star system, but the movement of additional matter in the DM density spike increases the rate of energy loss. This phenomenon was measured recently in a couple of X-Ray binaries (Chan & Lee 2023). These DM density spikes, called Ultra-Compact Mini Halos (UCMHs) when referring to relics of galactic halos and SMBHs (Yang 2022), were originally thought to be a collection of Weakly Interacting Massive Particles (WIMPs) trapped around a primordial black hole, yet unable to lose enough angular momentum to pass the event horizon. The UCMH and black hole together are called DM halo black holes (DMHBHs) (Stelea et al. 2023). Since we propose that DM density spikes are produced with the formation of MACHOs that are not dwarf stars (Ireland 2025), and that these vary with the size of their host, the term UCMH will be used for simplicity.

The need for a weak-force mediator in the Strong CP-Problem of reorganizing quarks, and the wave-like nature of galactic DM halos (Amruth et al. 2023), suggest axions are a better exotic particle candidate than WIMPs, since axions are Weakly Interacting Slender Particles (WISPs). The virial mass of an UCMH is approximately twice that of the host black hole (Chan & Lee 2023), so a large portion of the exotic DM could be bound tightly to MACHOs. Therefore, the overall number of MACHOs needed to explain all of the DM mass can be decreased by up to two-thirds (Chan & Lee 2023). As the black holes get bigger, the mass of the UCMH may decrease relative to the host black hole due to excitement from the environment. However, the exact relationship, and the average fraction of exotic DM, need further study.

The "Diversity Problem" (Nadler et al. 2023) is the discrepancy between observed galaxies and the common morphologies of similar-mass galaxies predicted by Cold DM and modern models of galactic evolution. SMBHs are considered ultra-Self-Interacting Dark Matter (uSIDM) (Pollack et al. 2015), and it has been shown that velocity-dependent Self-Interacting Dark Matter (SIDM) helps to solve the Diversity Problem (Nadler et al. 2023).

Evidence that DM is not simply collisionless has been gaining credence over the last decade (Nadler et al. 2023; Valdarnini 2024). In fact, the possible interaction of DM with regular matter that is not only gravitational has been proposed before (Almeida et al. 2024; Barkana 2018). The combination of a MACHO, with an axion halo in what is likely to be a DMHBH, seems to be the best candidate for explaining almost the entirety of DM. Henceforth, we will refer to this candidate as a WilloWISP. Distinctly different from Friedmons (Dokuchaev et al. 2014), also called "atomic DM" (Roy et al. 2023) or "dark atoms," a WilloWISP can be larger than microscopic; has a majority of WISPs in the UCMH; and it has no native leptons unless it is small enough to no longer support an event horizon and, therefore, has a possible nucleon shell. The details within a WilloWISP are discussed further in section 5.

## 2.    The Possible Size Range of Quark Stars

Since the pressure fluctuations and convection within the quarkonium in the core of a cooling quark star (Gómez-Bañón et al. 2024) might force some quarks together preferentially over others, collections of various sizes might form in suspension. For example, tetraquarks and pentaquarks (Sirunyan et al. 2022) ought to precipitate into quasi-crystals much like the diamonds in the center of the earth, or the metal crystals thought to form in the mantle of cooling white dwarf stars (Bédard et al. 2024). The exact density of a quark star is unknown, as well as that of quarkonium, but approximations can be made to glean the size range of such stars.

In the following discussion, we estimate the mass at which a quark star would be enshrouded in an event horizon. The most interesting case of the quarkonium equation of state requires that individual quarks be separated, and vibrate, in space (Ferrer 2017; Kanakubo et al. 2022). To analyze the structure, one can assume that the center starts with no mass in a meson condensate, and then it grows over time. The rate of this growth is unknown. However, if there is a bottleneck of some kind, i.e. reorganizing the quarks into acceptable meson pairs, then a neutron star could become reduced to the quarkonium proposed to be growing in their cores (Annala 2000; Bhattacharya 2023) without an event horizon.

The measurement of the volume of individual quarks is essentially impossible due to the nature of the strong force and pair production through flux tubes. However, attempts to estimate their contact sizes have been made before, which tend to be less than $10^{-19}$ meters, while nucleons are on the order of $10^{-15}$ meters, known as a fermi, or a femtometer (Khachatryan et al. 2015). The least dense of these quarks ought to set an upper limit on a quark star's radius, $R_*$, assuming the densities, $D_q$, are near-absolute-zero temperatures (Eq. 1-3; Table 1). We can then compare this estimated radius to the Schwarzschild radius of the same mass to estimate the size of a quark star that does not have an event horizon. The quarks with the largest rest mass might have the largest possible contact sizes, and strangeness keeps the largest flavors from forming (Alford et al. 2006), so the charm quark's mass is used as an overestimate with a conservative upper bound on the occupational volume. This demonstrates that the critical mass at which a quark star is naked, $M_*$, must be much less than a solar mass.

$$V \equiv \frac{4\pi}{3}r^3 \qquad \text{(Volume of a Sphere)} \qquad (1)$$

$$R_* \equiv \frac{2GM}{c^2} \qquad \text{(Schwarzschild Radius)} \qquad (2)$$

$$D_q \equiv \frac{M_q}{V_q}Q_{PF} \qquad \text{(Quark Quasi-Crystal Density)} \qquad (3)$$

$$Q_{PF} \equiv 0.74 \text{ (HCP: Maximum Quark Packing Factor)}$$

$$\left(1.78 \cdot 10^{-30}\,\frac{kg\,c^2}{MeV}\right) \text{ For mass unit conversion}$$

**Table 1**

Estimating the Density Range of Quark Stars

|  | Up | Down | Strange | Charm |
|---|---|---|---|---|
| **Radius (M)** | ~1.00E-21 | ~5.00E-20 | ~1.00E-19 | ~1.00E-16 |
| **Mass (MeV/c²)** | ~2.20 | ~4.79 | ~95.0 | ~1,270 |
| **Density (kg/m³)** | ~9.35E32 | ~2.00E28 | ~4.04E28 | ~2.26E21 |

**Notes.** Average densities of pure-flavor quarkonium from their rest masses and over-estimated contact sizes from proton-proton collisions. Assigning order of magnitude estimates to each flavor demonstrates that only the lightest flavors are capable of supporting a quark star of significant mass with no event horizon, as seen in Table 2.

**References.** Khachatryan et al. 2015; Blaschke et al. 2001; Rohlf 1994

As seen in Table 1, the minimum density of very cold down quarkonium (downonium) is ~$2 \cdot 10^{28}$ kg/m³, assuming no meson condensates. Comparing this with the density of a nucleon superfluid, ~$5 \cdot 10^{17}$ kg/m³ (Blaschke et al. 2001), in a neutron star of 11 km radius and 1.4 $M_\odot$, one can estimate the maximum size of a quark star in Table 2. The quarks in a neutron (two down and one up quark) imply that newly-formed quarkonium in a neutron star will contain a majority of down quarks, about a third of the whole in up quarks, and a handful in others. Matter containing strange quarks is more stable than nucleons (Alford et al. 2006), so the known runaway reaction to spread strangeness on contact with other matter could be free to drive the growth of the core BECs in a quark star. This would also convert most of the deconfined up quarks into strange quarks (Alford et al. 2006) with increasing probability as one approaches the quark star's center.

Now, using the definition of average density, $D_q$:

$$V_*(R_*) = \frac{M_*}{D_q} = \frac{4\pi}{3}\left[\frac{2GM_*}{c^2}\right]^3 \qquad (4)$$

Substitute Eq.3 into Eq.4 to obtain:

$$M_* = Q_{PF}\,M_q\left(\frac{R_*}{R_q}\right)^3 = Q_{PF}\,M_q\left(\frac{2GM_*}{R_q c^2}\right)^3 \qquad (5)$$

solve for $M_*$

**Table 2**

Examples from the Range of Possible Quark Star Sizes

| Quark Density | $M_* = \dfrac{c^3}{\sqrt{Q_{PF}\,M_q}}\left[\dfrac{R_q}{2G}\right]^{\frac{3}{2}}$ ($M_\odot$) | $R_* = \dfrac{2GM_*}{c^2}$ (meters) |
|---|---|---|
| Up | ~3.43E-7 | ~1.01E-3 |
| Down | ~9.00E-5 | ~1.00 |
| Strange | ~3.00E-5 | ~1.40E-1 |
| Charm | ~0.200 | ~6.30E-2 |
| Neutronium | ~6.00 | ~1.77E4 |

**Notes.** Maximum masses and radii of naked quark stars (no event horizons) if entirely comprised of a single flavor, and if the contact size is increasingly over-estimated with larger flavors. These are compared to Neutronium (neutron star material).

The density of the other quark flavors and the possibility of meson condensation imply that the star's average density is largely underestimated by assuming a majority of quarks are of the down or strange flavors. However, the extreme pressure and rotation at the center of a neutron star could maintain temperatures great enough to support a quark star of slightly more mass than the range of critical masses found here suggest.

## 3.    A New Dark Growth Channel for Large Black Holes

Regardless of the exact density of naked quark stars, the Schwarzschild Radius quickly overtakes the surface radius of one that approaches the Chandrasekhar Limit (~1.4 M$_\odot$) (Chandrasekhar 1984). As a neutron star cools, the final mass determines if the product is a quark star or a black hole. Boson stars composed of mesons are far more dense than quarkonium, by definition. Therefore, we suggest that a new growth channel for black holes be added to the existing list, namely: Supernovae, coalescence, primordial direct collapse (Clesse & Garcia-Bellido 2017; Hooper et al. 2024; Hütsi et al. 2023; Su, Li, & Feng 2023), dark energy (Farrah et al. 2023), and large star mergers (Fujii 2024).

This new channel could also produce exotic stars if the pulsar loses enough mass to avoid an event horizon while maintaining enough pressure in the core to harbor quarkonium (Eq. 1-5; Table 2). A neutron star left alone to cool off is supposed to be extremely stable, but an exotic matter core (Annala 2000; Bhattacharya 2023) allows a pathway for transmutation (Yakovlev et al. 2004). If enough mass is lost through neutrino emission (Blaschke et al. 2001), or axion production (Noordhuis et al. 2023), self-gravity might become weak enough to shed the nucleon mantle (Cheng & Dai 1996), which may have been observed in GRB 240529A (Tian et al. 2025).

As a pulsar ages, entropy increasing in the universe ought to be reflected in the cooling of the core, which then gradually causes collapse into quarkonium, and possibly further still into a set of BECs. This process is akin to evaporative cooling, combined with the laser-cooled chambers used for creating BECs in the laboratory (Blaschke et al. 2001). Except in a pulsar, the lasers are replaced by gravity, and the momentum of photons, axions, and leptons (mostly neutrinos) leaving the star's surface provide the evaporative effect (Blaschke et al. 2001). The core falling further into a highly ordered state consumes any excess nucleon mass in the mantle of the neutron star in what should be called a quark drip to mirror the established neutron drip (Chandrasekhar 1984; Blaschke et al. 2001).

## 4.    Quantum Black Holes as Large Exotic Stars

Einstein's General Relativity predicts an inflection of the curvature of spacetime at the event horizon, such that time-like paths become space-like and vice versa (Ohanian & Ruffini 1994). Making sense of this kind of motion is difficult, and assuming this is the case, it implies that space around a given mass inside a black hole is being dragged directly to the center. While this seemed to imply the existence of a singularity before, a more reasonable conclusion is that electrodynamics is quantum-mechanically forbidden within an event horizon.

Any quantum of matter passing the event horizon with an electric charge would mean the black hole necessarily gains a quantum number for net charge. While it is remotely possible that black holes are able to have net charge (Reissner-Nordström and Kerr-Newman), their stability is well known to have problems due to the extreme preferential attraction for the opposite charge (Stelea et al. 2023). Considering the fact that the black hole likely must be neutral in electric charge (Stelea et al. 2023), and that individual quarks only have fractional charges, the event horizon of a black hole may instead be an EM phase transition.

If true, then the event horizon's hypothesized firewall (Dai 2020), and possibly hair (Stelea et al. 2023), may instead be an extremely textured and warped boiling of spacetime coupled with the segregation of leptons and neutral quark configurations. This is supported by the possible measurement of "Electron-positron jets associated with the quasar 3C279 (Wardle et al. 1998)," and the "Evidence for quark-matter cores in massive neutron stars (Annala 2000)" from the simulation of the star's speed of sound (Mroczek et al. 2023). As conformal transformations are important in quantum field theory, "if the conformal bound $c_s^2 \leq 1/3$ is not strongly violated, [where $c_s$ is the speed of sound,] massive neutron stars [above ~1.4 $M_\odot$] are predicted to have sizable quark-matter cores" (Annala 2000). A cross-section of a neutron star with an exotic core can be seen in (Yunes et al. 2022).

The energy needed to rearrange quarks into neutral configurations, and drive the leptons away in jets, can easily be powered by the plunging region recently discovered around a black hole named MAXI J1 820+070 (Mummery 2024). The static charges and magnetic fields formed from friction easily tear charges away from each other (separating leptons from nucleons) and funnel them toward the poles (McKinney 2006). If a high density of photons builds up near the event horizon, then they can eventually combine in the extreme magnetic fields to form additional axions in the UCMH, since the opposite reaction is supposed to exist (Noordhuis et al. 2024). Such a process would explain the lack of superradiance in spinning black holes (Du et al. 2022), because the ergosphere should excite photons with appropriate trajectories into high enough energy levels to escape.

A rho-meson (up and down quark pairs) condensate should already be seeded in heavy neutron stars (Mallick et al. 2014) within the quarkonium core. This, plus the fact that a meson condensate would need to have an even number of quarks means that the remaining fermionic matter would need to be expelled, or orbit, the exotic core.

BECs are typically made in the laboratory from one type of boson (Blaschke et al. 2001), so the multitude of different mesons that are possible might have a hierarchical structure in orbitals much like the electrons around an atomic nucleus. While the detailed structure of the meson core will need to be refined in future work, a single strange meson condensate, Phi (strange and anti-strange quark pairs), is most likely the remaining structure within an event horizon because it avoids the unique quantum number for strangeness.

If the quarkonium is entirely converted to one, or a set of, meson BECs and an axion UCMH, yet there is no event horizon, then the object is similar to what is currently called a boson star (Brito et al. 2016; Bustillo et al. 2021; Chavanis & Harko 2012; Olivares et al. 2023; Pombo & Saltas 2023). Ultimately, boson stars could replace the need for a singularity in a black hole by filling the interior of the event horizon with a singular BEC wavefunction.

The idea that black holes are not simply singularities is not new, and the implications from a relatively new hypothesis concerning "frozen stars" (Brustein et al. 2024) is uncannily similar to the internal structure presented here. Frozen stars were first proposed to be high-metallicity stars that asymptotically appear similar to a black hole as a result of the heat death of the universe (Brustein et al. 2024). The lack of singularity and event horizon in frozen stars is now attractive as a potential solution to the infinities and paradoxes of the modern black hole model.

## 5.    The WilloWISP Model of the Black Hole Information Paradox

The interior of a black hole may have many possible structures in addition to, or instead of, the previously proposed singularities and their extensions (Siemonsen 2024; Azuma, Endo, & Koikawa 1991). Therefore, one can deduce a series of hypothetical structures from our WilloWISP model for the core of a black hole, and the interior of a quark star, which are based in quantum mechanics (Fig. 1). The approximate spherical volume calculated from the average density of hexagonally closely packed quarks marks the maximum surface radius of any quark quasi-crystals (Eq. 1-5) surrounding the BEC in the center of a quark star. Vacancies in the quasi-crystal closest to the center would be the result of a gluon vortex virtually testing every possibility of quark pairs that can be absorbed into the central meson condensate.

Moving away from the center, the quasi-crystalline structure of the outer core of a WilloWISP gives way to quarkonium (Fig. 1). The crust of an exposed quark star is likely a nucleon superfluid (Blaschke et al. 2001) like a neutron star has a crust of atomic nuclei. However, any fermionic matter should not exist inside a black hole, where the event horizon is well beyond the estimated quarkonium surface of a quark star with similar mass (Eq. 1-5).

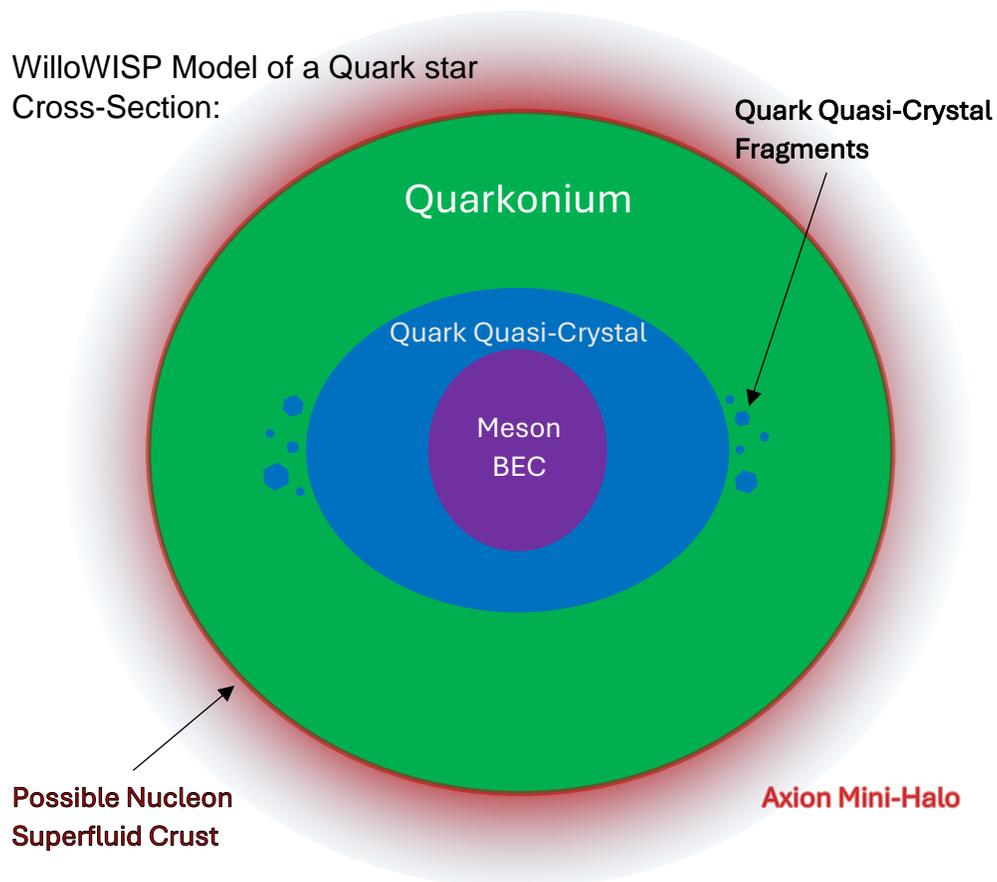

**Figure 1.** Internal structure of a quark star in an UCMH of predominantly axions, which provides the WISP in the name WilloWISP. The interior of a black hole can be deduced from this, which could be eventually exposed as a spinning boson star, if it remains stable. Therefore, a black hole is likely a strange meson BEC that interacts with matter and spacetime quantum-mechanically through the event horizon, as it is a singular wavefunction.

The accretion disk, itself, warps the event horizon of a black hole slightly, by the fact that a massive ring's gravitational outward pull becomes stronger with distance from the central axis inside the ring. This essentially pinches the equator of the event horizon inward, while the poles bulge outward slightly, making somewhat of a subtle lemon shape, with one of the jets pointing out of what would be the stem (Fig. 2). Since there is a slight precession of the black hole's jet with its accretion disk, this shape becomes dynamically perturbed while in the act of feeding, much like the ripples in the surface of water falling down a drain. This phenomenon is reminiscent of the hypothesized holographic encoding on the surface of black holes, as well as the proposed hairs outside their supposed firewall (Dai 2020).

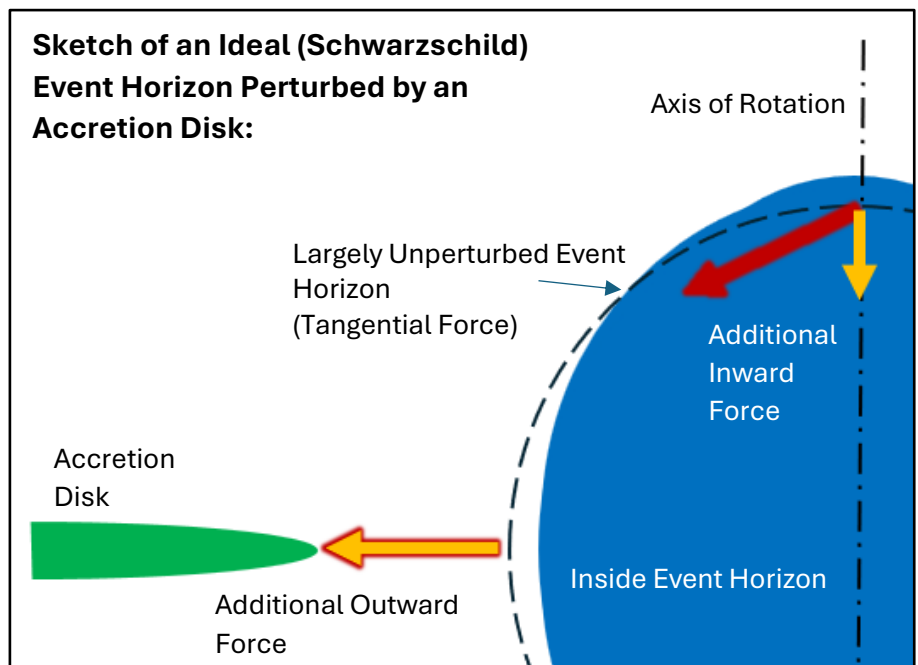

**Figure 2.** Event horizon (blue area's boundary) of a feeding black hole. If the black hole has an active accretion disk, then the perturbation of the event horizon (the original is marked by a dashed circle) should result in a pinched equator and protruded poles. Red arrows represent the force of the disk at the surface of the unperturbed event horizon. Orange arrows represent the component of the disk's force is perpendicular to the dashed circle. The effect is exaggerated for clarity, and not to scale.

Objects that are about to pass the event horizon do so in particle form, so that each fundamental mass is addressed individually. The single-dimension particle streams broken off from the two-dimensional inner edge of the accretion disk suggest fractal Lévy walks, which were recently found characteristic of pions in heavy-ion collisions (Kincses, Nagy, & Csanád 2025). Since the extreme tidal forces and simultaneity confine the brownian motion used to derive the familiar Schrodinger Wave Equation, a fractional quantum mechanics (Laskin 2000) is necessary to approach a solution to the quantum black hole (Jalalzadeh, da Silva, & Moniz 2021), which is often found to be of the soliton family (Ahmad et al. 2023; Azuma, Endo, & Koikawa 1991). However, this is a new field of study and, therefore, needs refinement before a treatment can be fully applied.

When a particle about to be consumed approaches a black hole's surface, the event horizon recedes to accept each one, since the local radius decreases due to the miniscule additional outward force. The quantum deformation and relaxation of the surface could send ripples into the cosmos via the axion UCMH, which would track at least a fraction of the information thought to be lost in this process.

Hence, a solution to The Black Hole Information Paradox could be deduced from the connection between the quark composition of the black hole's core and the history of the matter passing through the event horizon, the encoding mechanism of which can be partially seen in Figure 2. The well-known after-image of matter asymptotically approaching the event horizon allows the EM information to remain in normal spacetime, representing the organization that would need to take place in a quark star (Fig. 1).

The only information potentially escaping a black hole is in the form of Hawking Radiation, gravitational waves, and the accretion's jets. Therefore, these might reflect the information in the event horizon's deformation and the core's virtual arrangement of matter. As the black hole evaporates entirely through Hawking radiation, the information could be at least partially recovered by annihilation of the core's history of available surface states with entangled particle pairs (much like the game of Mahjong).

A WilloWISP that is small enough to be exposed to cosmic radiation without an event horizon, however, is likely to be unstable. The resulting explosion from decoherence of the BEC (Lukin et al. 2021) could partly explain the Gamma Ray Excess (GRE) in galactic centers (Goodenough & Hooper 2011).

## 6. Methods of WilloWISP Detection

A recent publication (Liu 2024) demonstrates that "gravitational-lasers" (called gasers, at least when artificial) (Zouros 1979) are possibly emitted by a black hole with an UCMH of axions (a WilloWISP), which is consistent with the fact that general relativity supports this type of "stimulated gravitational radiation". The theory suggests that instruments like LIGO-VIRGO-KAGRA, but more sensitive, could test the existence of axions in UCMHs by detecting these "gasers." The planned Laser Interferometer Space Antenna (LISA) is potentially sensitive enough (Colpi et al. 2024) to do this and, thereby, allows the deciphering of information once thought lost.

Microlensing of WilloWISPs produces a slightly different light curve than only a MACHO (Macedo et al. 2024). The UCMH extends the microlensing event time and softens the intensity curve, which should not be the case for dwarf stars. Therefore, a statistical sampling of light curves can estimate the population of WilloWISP-like MACHOs in a given stellar ensemble.

## 7.   A Source of Dark Matter: From Neutron Star aging to IMBHs

The formation rate of neutron stars in the Milky Way Galaxy is thought to be ~ 1 every 30 years, or about $10^9$ over its lifetime (Endal 1979; Chandrasekhar 1984). However, the rate is debatable, possibly being as high as once every four years (Endal 1979). Approximately one out of every 1000 stars are large enough to collapse into black holes instead of neutron stars via supernovae (Chandrasekhar 1984). Coincidentally, the larger masses of black holes compared to neutron stars, and their better ability to accumulate mass, estimates their population in The Milky Way Galaxy to be also about $10^9$ (Agol 2002). The number of IMBHs in the Milky Way Galaxy needs to only reach ~1% of these to explain all of DM's missing mass. The rates of MACHO coalescence, migration, and production from stars are largely unknown, but they can now be estimated by what would be needed to meet the mass budget of a given galaxy. If the equation of state for quarkonium and a meson condensate together can be found, then maybe the process of condensation can be compared to neutron star formation rates to better grasp the growth rates of the large black holes we see today.

Any neutron star collisions further accelerate black hole growth, but these would leave bright tracers, and the cross section is extremely small. The coalescence of black holes or exotic stars, however, would only be detected by gravitational waves. The fact that ~90% of all stars are likely to be main-sequence dwarf stars (ESA 1997), and that the neutron star population is more difficult to measure than expected in dense matter regions (Endal 1979), implies a leak in the mass budget of the stellar life cycle. If neutron stars are able to cool and transmute (Yakovlev et al. 2004), then the mass leak is explained, and a relatively large number of objects in the universe are continuously in the process of producing MACHOs of about a solar mass at most. Considering that these processes are likely to happen more frequently in dense matter regions due to high stellar density and prolific large stars, these small MACHOs could provide a fraction of the mass required to feed the IMBHs at the same time as their axion UCMHs coax other MACHOs together.

## 8.   Conclusion

A quarkonium core in neutron stars (Annala 2000) becoming quark stars with age could explain: the production of axion clouds around the poles of pulsars (Noordhuis et al. 2023); the difficulty in detecting neutron stars (Endal 1979); and the internal structure of a black hole. The axion connects quarkonium, black hole coalescence (and accretion), and the formation of astrophysical jets with quark degeneracy. The fact that axions can be traced throughout the aging process provides strong evidence that neutron stars and black holes are heavily correlated. Therefore, we suggest that MACHOs (excluding dwarf stars) can be described with our WilloWISP model, and that the evaporation of neutron stars ought to be added to the list of potential dark growth channels for black holes.

A means to produce exotic stars instead of black holes further lends dark mass to the quiescent feeding process of large black holes, even if it is a tiny fraction of the total DM. At the same time, this process explains the observed "paucity" of black holes less than ~5 $M_\odot$ (Özel et al. 2010). Since equations 1-5 assumed a nearly absolute zero temperature, and ignored quark-flavor mixing, the natural vibration of quarks and gluons slightly increases the maximum mass before a quark star is indistinguishable from a black hole (Macedo et al. 2024), which may be equivalent to a boson star with age.

The motion of many small WilloWISPs ought to accelerate coalescence rates of larger WilloWISPs without the need for galactic mergers or bright tracers, since smaller objects remove kinetic energy from larger objects, and UCMHs increase orbital decay rates. In this way, hierarchically clustered IMBHs could explain all of the DM with so few objects, with such small number density, that they could almost entirely hide from past microlensing event surveys (Rahvar 2005; Macedo et al. 2024) and CMB constraints (Coriano & Frampton 2021; Su, Li, & Feng 2023).

The loss of radius due to neutrino emission, cooling temperatures, and the jets, would increase any existing angular velocity, perhaps turning large neutron stars into pulsars over time without the need for a companion star (Blaschke et al. 2001). As it spins faster with age, the core collapse has a necessarily competing centrifugal force that increases the chances of the neutron star forming a transitional exotic star before becoming a tiny black hole. The quarks in an exotic star are quickly converted to the strange flavor, since strangeness spreads on contact in these conditions. Therefore, the process discussed here implies that black holes are a strange boson star behind the event horizon, which is included in our proposed WilloWISP model of MACHOs.

Since exotic DM seems to be highly localized around MACHOs, it is very possible that clusters of these WilloWISPs could be roaming the cosmos, quiescently contributing the majority of the DM's missing mass since the era of first stars (Cann et al. 2018; Hütsi et al. 2023). There has even been evidence of such a structure that has no other bright tracers besides the "Spur and Gap of the GD-1 Stellar Stream" in the Milky Way Galaxy (Bonaca et al. 2019; Zhang et al. 2025). If the recently discovered dark dwarf galaxies (Xu et al. 2023), Nube (Montes et al. 2023) and Ursa Major III (UNIONS-1) (Smith et al. 2024), are found to be mostly containing quiescent black holes, then the culprit of the spur and gap in GD-1 is likely a similar cluster of WilloWISPs.

## 9.    Acknowledgements

I thank my advisor, Prof. Neil F. Comins, for providing resources for this work, and for giving me the opportunity to acquire a doctorate. Partial financial support of this work for a short time by the Maine Space Grant Consortium was much appreciated. I am grateful to the external reader of my dissertation, Prof. Hai-Bo Yu, for providing specialized evaluation of my work that further encouraged me to share it. I would also like to thank my friends Dr. George P. Bernhardt IV, Prof. Tom Stone, and Dr. Sarah B. Rice for their support and detailed comments on this work. Furthermore, I would not have the advantages of a vast knowledge base at a young age, and a unique understanding of nature, without the life-long guidance from my parents, Prof. Scott D. Collins and Prof. Rosemary L. Smith.